\documentstyle[preprint,prl,aps]{revtex}
\begin{document}
\draft
\title{Weak levitation of 2D delocalized states in a magnetic field}
\author{T.~V.~Shahbazyan and M.~E.~Raikh}
\address{Department of Physics, University of Utah, Salt Lake City, UT
84112}
\maketitle
\begin{abstract}
The deviation of the energy position of a delocalized state from the
center of Landau level is studied in the framework of the
Chalker-Coddington model. It is demonstrated that introducing a weak
Landau level mixing results in a shift of the delocalized state up in
energy. The mechanism of a levitation is a
neighboring - Landau level - assisted resonant tunneling which ``shunts''
the saddle-points. The magnitude of levitation is shown to be
independent of the Landau level number.
\end{abstract}
\narrowtext

It is commonly accepted that electronic spectrum of a two-dimensional
disordered system in a strong magnetic field contains only a single
delocalized state per Landau level (LL). First conjectured by
Halperin \cite{hal82}, this conclusion was later drawn from the
consideration based on scaling ideas \cite{lev83,khm83} and confirmed
by numerical simulations in which the divergence of the localization
length with energy approaching to the position of the delocalized
state was studied \cite{aok85,huc90,huo92}.

To compromise this conjecture with the absence of delocalized states
at zero field ($B=0$) Khmelnitskii \cite{khm84} has suggested that with
decreasing $B$ the energy positions of delocalized states depart
gradually from the centers of Landau levels and float upwards to
reach an infinity as $B\rightarrow 0$. In order to obtain the positions of
delocalized states, $E_n(B)$, Khmelnitskii has pointed out that the
system of scaling equations \cite{lev83} for conductivities, $\sigma_{xx}$,
$\sigma_{xy}$ should be solved together with the boundary conditions
that at small scales those are given by their classical expressions:
$\sigma_{xx}^{cl}=\sigma_0(1+\omega_c^2\tau^2)^{-1}$,
$\sigma_{xy}^{cl}=\sigma_0\omega_c\tau(1+\omega_c^2\tau^2)^{-1}$,
where $\sigma_0$ is the Drude conductivity (proportional to the Fermi
energy $E_F$), $\omega_c$ is the cyclotron frequency and $\tau$ is the
elastic scattering time. Since the scaling equations are periodic in
$\sigma_{xy}$, it follows that $\sigma_{xx}(E_F)$ remains finite at
large scales each time when $\sigma_{xy}^{cl}=(n+1/2)e^2/2\pi\hbar$.
This gives $E_n=\hbar\omega_c(n+1/2)[1+(\omega_c\tau)^{-2}]$.
In support of the above picture Laughlin \cite{lau84} has presented an
argument that the delocalized states cannot disappear discontinuosly
at any finite magnetic field.

The levitation of the delocalized states according to \cite{khm84,lau84}
is sketched in Fig.~1. The above dependence $E_n(B)$,
treated as a dependence of $\rho_{xx}^0=\sigma_0^{-1}$ versus the
inverse filling factor is, in fact, an essential constituent of the global
phase diagram of the quantum Hall effect (QHE) introduced by Kivelson,
Lee, and Zhang \cite{kiv92} (namely, an integer QHE part of this diagram).
Recent experiments \cite{jia93,joh93,wan94,hug94}, in which
insulator - quantum Hall conductor - insulator transition was observed with
increasing magnetic field, were interpreted in terms of the global phase
diagram \cite{kiv92}. Using Fig.~1 the interpretation presented in
\cite{jia93,wan94,hug94} can be reformulated as follows. If the Fermi
level at zero field lies within some certain range, as shown in
Fig.~1,  then the horizontal line, $E_F(B)=const$, crosses the curve
$E_0(B)$ at two different values of $B$. This implies that for $B$ in
between these values the system is in the QHE regime
while for smaller or larger $B$ the system is in the insulating phase.
The latest experiment \cite{glo95}, carried out at very low
temperatures, allowed, in fact, a direct measurement of $E_0(B)$
dependence by tracing the positions of peaks in $\sigma_{xx}$,
measured as a function of gate voltage, at different magnetic fields.
The measured dependence $E_0(B)$ exhibits a minimum in the region of
$B$ where spin levels are not resolved.

Numerical simulations \cite{and84} for the case of a short-range
disorder indeed indicated an upward (from $E=\hbar\omega_c/2$) shift of
the energy position of the extended state when three lowest LL were
taken into account. However, no microscopic theory for such a
levitation has been developed so far. In the present paper we point
out that such a theory can be developed in the region where the
departure of the delocalized state from the center of the Landau band
is relatively small. We demonstrate that the microscopic mechanism for
the levitation is the resonant scattering of an electron with energy
near the center of LL by localized states from neighboring LL.

We consider the case of a smooth random potential and adopt the
network model by Chalker and Coddington \cite{cha88} which we
generalize in order to study the effects of neighboring LL. In this
model delocalization results from a quantum tunneling of an
electron through saddle-points of a smooth potential which are
connected by equipotential lines. The randomness of the potential is
included by assuming the phase acquired by electron traversing a link
(equipotential line) to be random. The probability amplitudes of
incoming waves $Z_1$ and $Z_2$ and outgoing waves $Z_3$ and $Z_4$
(see Fig.~2) are related as

\begin{eqnarray}\label{1}
\left(\begin{array}{c}
Z_1\\
Z_3
\end{array}\right)
={\bf M}
\left(\begin{array}{c}
Z_4\\
Z_2
\end{array}\right),
\end{eqnarray}
where the matrix ${\bf M}$ can be decomposed into the product

\begin{eqnarray}\label{2}
{\bf M}=
\left(\begin{array}{cc}
e^{i\phi_1} & 0\\
0 & e^{i\phi_2}
\end{array}\right)
\left(\begin{array}{cc}
\cosh\theta & \sinh\theta\\
\sinh\theta & \cosh\theta
\end{array}\right)
\left(\begin{array}{cc}
e^{i\phi_3} & 0\\
0 & e^{i\phi_4}
\end{array}\right),
\end{eqnarray}
with parameter $\theta$ characterizing the tunneling and $\phi_i$
being gauge phases. For the potential  expanded near the saddle
point, $V(x,y)=V_0-m\Omega_x^2 x^2/2+m\Omega_y^2 y^2/2$, $m$ being the
electron mass, this parameter can be presented as \cite{fer87}

\begin{eqnarray}\label{3}
\sinh\theta=\exp\Biggl({E-V_0\over \gamma}\Biggr),
{}~~~~~\gamma=\hbar\Omega_x\Omega_y/\pi\omega_c,
\end{eqnarray}
where $E$ is energy of electron measured from the center of LL. Since
the saddle-point heights, $V_0$, are distributed {\em symmetrically}
around the value $V_0=0$ [corresponding to
$\theta=\theta_c\equiv\ln(1+\sqrt{2})$], the delocalized states occurs
at zero energy (Chalker and Coddington assumed $V_0=0$ at all saddle
points; later simulations \cite{lee93} with randomly distributed
$V_0$ demonstrated the universality of the critical exponent).

Assume now that the energy is close to the center of $n=0$ LL and study
the change in the structure of electron states caused by $n=1$ LL.
The relevant $n=1$ LL states are those with energies close to the
center of $n=0$ LL. The equipotentials
corresponding to these states are depicted schematically in Fig.~2 by
dashed loops. The prime role of these equipotentials is to shift the
maximum of the density of states up from $E=0$. However, such a shift
does not affect the position of delocalized state unless a coupling
between the equipotentials of $n=0$ and $n=1$ LL is introduced. This
coupling is illustrated in Fig.~2a. The line connecting the
equipotentials stands for a scattering matrix ${\bf S}$ defined as

\begin{eqnarray}\label{4}
\left(\begin{array}{c}
Z\\
\tilde{Z}
\end{array}\right)
={\bf S}
\left(\begin{array}{c}
Z'\\
\tilde{Z}'
\end{array}\right),
\end{eqnarray}
which has a decomposition

\begin{eqnarray}\label{5}
{\bf S}=
\left(\begin{array}{cc}
e^{i\phi} & 0\\
0 & e^{i\tilde{\phi}}
\end{array}\right)
\left(\begin{array}{cc}
\cos\alpha & -\sin\alpha\\
\sin\alpha & \cos\alpha
\end{array}\right)
\left(\begin{array}{cc}
e^{i\phi'} & 0\\
0 & e^{i\tilde{\phi}'}
\end{array}\right),
\end{eqnarray}
Such a form of ${\bf S}$ provides the conservation of flux
$|Z|^2+|\tilde{Z}|^2=|Z'|^2+|\tilde{Z}'|^2$. The coupling strength is
characterized by an angle $\alpha$ (gauge phases in matrix ${\bf S}$,
as well as in ${\bf M}$, can be absorbed into $Z$'s).
It is easy to see that the only effect of such a coupling to a loop
is just a phase shift between amplitudes $Z$ and $Z'$. Indeed,
$\tilde{Z}$ and $\tilde{Z}'$ differ by a phase factor acquired by
an electron traversing the loop: $\tilde{Z}'=\tilde{Z} e^{i\varphi}$.
Then (\ref{4}) yields $Z=Z' e^{i\delta}$, with

\begin{eqnarray}\label{6}
\tan\delta(E)={\sin\varphi\sin^2\alpha \over
\cos\varphi(1+\cos^2\alpha)-2\cos\alpha}.
\end{eqnarray}
The energy dependence of the scattering phase $\delta$ is determined by
the energy dependence of $\varphi$. For a small coupling the resonances
occur at $\varphi=\pm\alpha^2/2+2\pi p$ ($p$ is an integer) the resonance
width being also $\alpha^2/2$. However, these resonances are of no
importance since the phase $\delta$ can also be absorbed into the
random phase on the link.

The situation is completely different when a loop occurs in the
vicinity of the saddle-point so that it is coupled to the both
incoming and outgoing links. It is important to note that such a loop
is not coupled directly to the saddle-point because the gradient of
potential there is precisely zero. The tunneling between the loop and the
links occurs most favorably at a certain distance from the saddle
point, as illustrated in Fig.~2b.
Since the characteristic tunneling distance at the saddle-point is quite
small (typically, the magnetic length), we neglect the effect of the
loop on tunneling transparency at the saddle-point.

Below we demonstrate that the effect of such loops is to shift the
position of delocalized state up. The crucial observation is that the
saddle-point with the loops attached to the links can be viewed as
some modified saddle-point (see Fig.~2b) and, thus, characterized by a
matrix ${\bf M}'$ with the same unitarity properties

\begin{eqnarray}\label{7}
\left(\begin{array}{c}
Z'_1\\
Z'_3
\end{array}\right)
={\bf M}'
\left(\begin{array}{c}
Z'_4\\
Z'_2
\end{array}\right),
{}~~~
{\bf M}'=
\left(\begin{array}{cc}
e^{i\phi'_1} & 0\\
0 & e^{i\phi'_2}
\end{array}\right)
\left(\begin{array}{cc}
\cosh\theta' & \sinh\theta'\\
\sinh\theta' & \cosh\theta'
\end{array}\right)
\left(\begin{array}{cc}
e^{i\phi'_3} & 0\\
0 & e^{i\phi'_4}
\end{array}\right),
\end{eqnarray}
The new parameter $\theta'$ can be
expressed via $\theta$ and the elements of scatterig matrices
${\bf S}_i$ (see Fig.~2b) using the following equations

\begin{eqnarray}\label{8}
\left(\begin{array}{c}
Z'_2\\
Z_5
\end{array}\right)
={\bf S}_2
\left(\begin{array}{c}
Z_2\\
Z_6
\end{array}\right),
{}~
\left(\begin{array}{c}
Z_3\\
Z_6
\end{array}\right)
={\bf S}_3
\left(\begin{array}{c}
Z'_3\\
Z_5
\end{array}\right),
{}~
\left(\begin{array}{c}
Z'_1\\
Z_8
\end{array}\right)
={\bf S}_1
\left(\begin{array}{c}
Z_1\\
Z_7
\end{array}\right),
{}~
\left(\begin{array}{c}
Z_4\\
Z_7
\end{array}\right)
={\bf S}_4
\left(\begin{array}{c}
Z'_4\\
Z_8
\end{array}\right),
\end{eqnarray}
Consider for simplicity
a case when all $\alpha_i=\alpha$ and there is only one loop (e.g.,
${\bf S}_1={\bf S}_4={\bf 1}$). Then the solution of (\ref{1}), (\ref{7})
and (\ref{8}) reads

\begin{eqnarray}\label{9}
\sinh^2\theta'={2\sinh^2\theta\cos^2\alpha(1-\cos\varphi) \over
[\sin\varphi - \sin^2\alpha\cosh\theta\sin(\psi-\varphi)]^2 +
[\cos\varphi-\cos^2\alpha+\sin^2\alpha\cosh\theta\cos(\psi-\varphi)]^2}.
\end{eqnarray}
Here $\varphi$ [the same as in (\ref{6})] is the phase acquired
by electron traversing the loop while $\psi$ is the phase acquired on
the contour $Z_2\rightarrow Z_3 \rightarrow Z_5$
(see Fig.~2b).

The conversion of the parameter $\theta$ into $\theta'$ can be viewed
as an effective change in the height of
the saddle-point, $\delta V_0$, caused by $n=1$ LL:

\begin{eqnarray}\label{10}
\delta V_0=-\gamma\ln\Biggl({\sinh\theta'\over\sinh\theta}\Biggr),
\end{eqnarray}
The distribution function of $\delta V_0$, which comes from averaging
over the random phases $\varphi$ and $\psi$, is plotted in Fig.~3 for
different values of $\alpha$. The narrow peaks at small $\delta V_0$
originate from the values of $\varphi$ that are not too close to
$2\pi p$. Then it is easy to see from (\ref{9}) that for such
$\varphi$ and $\alpha\ll 1$
we have $\theta'\simeq\theta$ and, consequently, $\delta V_0$ is small.
At the same time, the tails of the distribution come from the
resonances $\varphi\simeq 2\pi p$. For such $\varphi$ we may have both
$\theta'>\theta$ and $\theta'<\theta$. Important is, however, that the
tails are {\em asymmetric}, i.e., the distribution function falls off
slower towards large $\delta V_0$. As a result, the average
$\overline{\delta V_0}$ appears to be positive. For small $\alpha$
this average can be easily calculated analytically:

\begin{eqnarray}\label{11}
\overline{\delta V_0}(\theta)=
{\alpha^2\gamma\over\pi}[\sinh\theta+\arcsin(1/\cosh\theta)],
\end{eqnarray}
where $\theta$ is related to $V_0$ by Eq.~(\ref{3}).
We see that $\overline{\delta V_0}(\theta)$ is proportional to the
width of the resonance, $\alpha^2$, since it is determined by resonant
loops. Since $\overline{\delta V_0}(\theta)$ is finite, the average
saddle-point height, $\overline{V_0}$, moves from $\overline{V_0}=0$
to a {\em finite} value
$\overline{V_0}=\langle\overline{\delta V_0}(\theta)\rangle_{V_0}$,
where $\langle ~ \rangle_{V_0}$ stands for averaging over $V_0$. The
values of $V_0$, relevant for delocalization, are of the order of
$\gamma$, the relevant $E-V_0$ being also of the order of $\gamma$, so
that the relevant $\theta$ in (\ref{11}) is $\theta\sim 1$. This leads
to the following estimate for the energy shift of the delocalized
state:

\begin{eqnarray}\label{12}
\delta E_0\sim\langle\overline{V_0}\rangle_{\alpha}
\sim\overline{\alpha^2}\gamma,
\end{eqnarray}
where $\overline{\alpha^2}$ is the coupling strength averaged over the
loops. It can be seen that the shift is much smaller than a typical
saddle-point height: $\delta E_0/\gamma \sim \overline{\alpha^2}\ll 1$.

Physically, the levitation (positive $\delta E_0$) originates from the loops
providing a direct transmission between links \cite{jai88}, bypassing
the saddle-point. In fact, some loops cause an opposite effect: by
acquiring an appropriate phase $\psi$ on the
contour $Z_2\rightarrow Z_3 \rightarrow Z_5$ the electron can be
better transmitted to the other side of the saddle-point. However,
this effect appears to be smaller than the effect of
bypassing. Thus, in order to compensate the leakage of electrons to
the opposite link via the loop, the energy of the delocalized state
raises up.

In the derivation of (\ref{11}) we neglected the effect of the loop
$Z_7\rightarrow Z_8$ at the other side of the saddle-point (see
Fig.~2b). It is easy to see, however, that the two loops cause
essentially an additive effect. Indeed, an electron, traversing the link
$Z'_1$, can also bypass the saddle-point due to a resonant transmission
to the opposite link $Z'_4$.

The key-point of the above consideration was that the loops can occur
only in directions of the descend of a saddle-point potential
(valleys), i.e., to the right and to the left from the saddle-point in
Fig.~2b. This is the case when the effect of $n=1$ LL on $n=0$
delocalized state is studied. For $n=1$ delocalized state the
situation is more complicated: the loops from $n=2$ LL occur in the
directions to the valleys while the loops from $n=0$ LL occur in the
directions to the hills. The latter loops cause an opposite trend,
pulling the delocalized state down. The resulting shift can be
presented as $\delta E_1 \sim \gamma
(\overline{\alpha_{12}^2}-\overline{\alpha_{10}^2})$. Note that the
coupling, $\overline{\alpha_{12}^2}$, of $n=1$ to $n=2$ LL is stronger
than the coupling, $\overline{\alpha_{10}^2}$, of $n=1$ to $n=0$ LL due
to a larger size of the $n=2$ wave function. Thus, $\delta E_1$ is
also positive. Another effect which we have neglected is coupling of
the $n=0$ links via the loops corresponding to $n=0$ equipotentials.
Since these loops occur with equal probability in {\em all} directions
from  the saddle-point, it is obvious that positive and negative
contributions to $\delta E_0$ from  these loops cancel each other out.

The effective coupling constants $\overline{\alpha_{n,n+1}^2}$ are
small if the LL width is small, i.e., $W\ll \hbar\omega_c$. But even
if $W\gg \hbar\omega_c$, the coupling can be still small since it is
determined by tunneling between equipotentials. A crude estimate for
a tunneling amplitude in this case, $W\gg \hbar\omega_c$, can be
obtained as follows. The spacial distance between equipotentials
is of the order of $s=\hbar\omega_c/({W\over R_c})$, $R_c$ being the
correlation radius of a smooth potential. This corresponds to the
momentum transfer $q=s/l^2$, $l$ being the magnetic length. The
coupling is efficient only if $q<R_c^{-1}$. Otherwise the smooth
potential will not be able to provide the necessary momentum, and the
tunneling amplitude would be exponentially small. The latter condition,
$qR_c\lesssim 1$, can be rewritten as

\begin{eqnarray}\label{13}
\hbar\omega_c \lesssim
\Biggl[W\biggl({\hbar^2\over mR_c^2}\biggr)\Biggr]^{1/2}
\equiv\hbar\omega_c^{*}.
\end{eqnarray}
Eq.~(\ref{13}) requires that the magnetic field should not be too strong;
however, the quasiclassical picture of electron motion along
equipotentials still
applies at $\omega_c\sim\omega_c^{*}$: for such $\omega_c$ we
have $l/R_c \sim [({\hbar^2\over mR_c^2})/W]^{1/4}\ll 1$,
which is just the condition that the potential is smooth.
On the other hand, the use of the Chalker-Coddington model is
justified only if the saddle-point ``discriminates'' neighboring LL,
i.e., $\hbar\omega \gtrsim \gamma$.
Remarkably, the two conditions, $\omega_c\sim\omega_c^{*}$ and
$\hbar\omega_c\sim\gamma$, coinside. Indeed, following Polyakov and
Shklovskii \cite{pol94}, the parameter $\gamma$ in (\ref{3}) can
be estimated as $\gamma \sim
\hbar({W\over mR_c^2})/\omega_c\sim\hbar\omega_c^{*2}/\omega_c$. Thus,
the Chalker-Coddington model is inadequate for
$\omega_c \ll \omega_c^{*}$, when the saddle-point at
$\hbar\omega_c(n+1/2)$ becomes transparent also for several
neighboring to $n$th LL; for such fields one might expect a strong
levitation. Our consideration applies for magnetic fields
$\omega_c > \omega_c^{*}$, when the levitation is weak.

Let us briefly discuss the dependence of the coupling on LL number
$n$. This dependence is determined by the overlap of $n$th and
$(n+1)$th Landau wave-functions shifted in space by a distance $s$.
This overlap is known to be proportional to
$e^{-u/2}L_n^1(u)[u/(n+1)]^{1/2}$, where $u=s^2/2l^2$ and $L_n^1$ is
the Laguerre polynomial. If the ratio
$\omega_c/\omega_c^{*}=(qR_c)^{1/2}=(R_cs/l^2)^{1/2}$ is not too large
(the coupling is not exponentially small) we have $s<l$. Then it is
easy to see that
$\overline{\alpha_{n,n+1}^2}\propto [L_n^1(0)]^2/n\propto n$. As a
result, the levitation $\delta E_n$, which is proportional to
$\overline{\alpha_{n,n+1}^2}-\overline{\alpha_{n,n-1}^2}$, does not
depend on $n$.

In conclusion, we have traced how the LL mixing gives rise to a shift
of the energy of the delocalized state up from the position
$\hbar\omega_c(n+1/2)$. In a smooth potential the shift may be
relatively small even when the peaks in the density of states
corresponding to different LL are not well-resolved [it can be seen
from (\ref{13}) that
$\hbar\omega_c^{*}/W=[({\hbar^2\over mR_c^2})/W]^{1/2}\ll 1$].
Surprisingly, the
simulation \cite{and84} for the {\em white-noise} random potential,
where one might expect much stronger coupling between LL,
indicated that for $\hbar\omega_c\simeq W$ the levitation is still very weak.

The picture considered above was a single-particle picture of
delocalization. In other words, we have completely neglected the
effects of screening caused by electron-electron interactions. At the
same time, it is known that in a smooth potential the electron-electron
interactions affect drastically the distribution of electrons within
the plane. In fact, when the potential is very smooth the
equipotentials, we have dealt with, are
separated by incompressible strips \cite{efr88}. However, as it was
discussed, a noticeable levitation occurs only when the field is not
too strong, so that the distance between equipotentials is less than the
magnetic length. Clearly, in this case the macroscopic description of
\cite{efr88} does not apply and a single-particle picture should be
valid.

\acknowledgments

Authors acknowledge the discussions with D.~E. Khmelnitskii and
D.~G. Polyakov on subjects related to this work.

\begin{figure}
\caption{Schematic picture of the levitation of delocalized states
according to [7,8]. The straight lines correspond to $n=0$ and
$n=1$ Landau levels.}
\end{figure}

\begin{figure}
\caption{Sketch of the saddle-point. The full lines represent
equipotentials corresponding to $n=0$ LL. The dashed circles are
equipotentials of $n=1$ LL. (a) Scattering of an electron by isolated
loop (${\bf S}$ stands for the scattering matrix). (b) Passage of electron
through the saddle-point in the presence of loops connecting the links.}
\end{figure}

\begin{figure}
\caption{The distribution function of $\delta V_0$
is shown at $E=V_0$ ($\theta=\theta_c$)
for $\alpha^2=0.1$ (solid line), $\alpha^2=0.2$ (long-dashed line), and
$\alpha^2=0.3$ (dashed line).}
\end{figure}

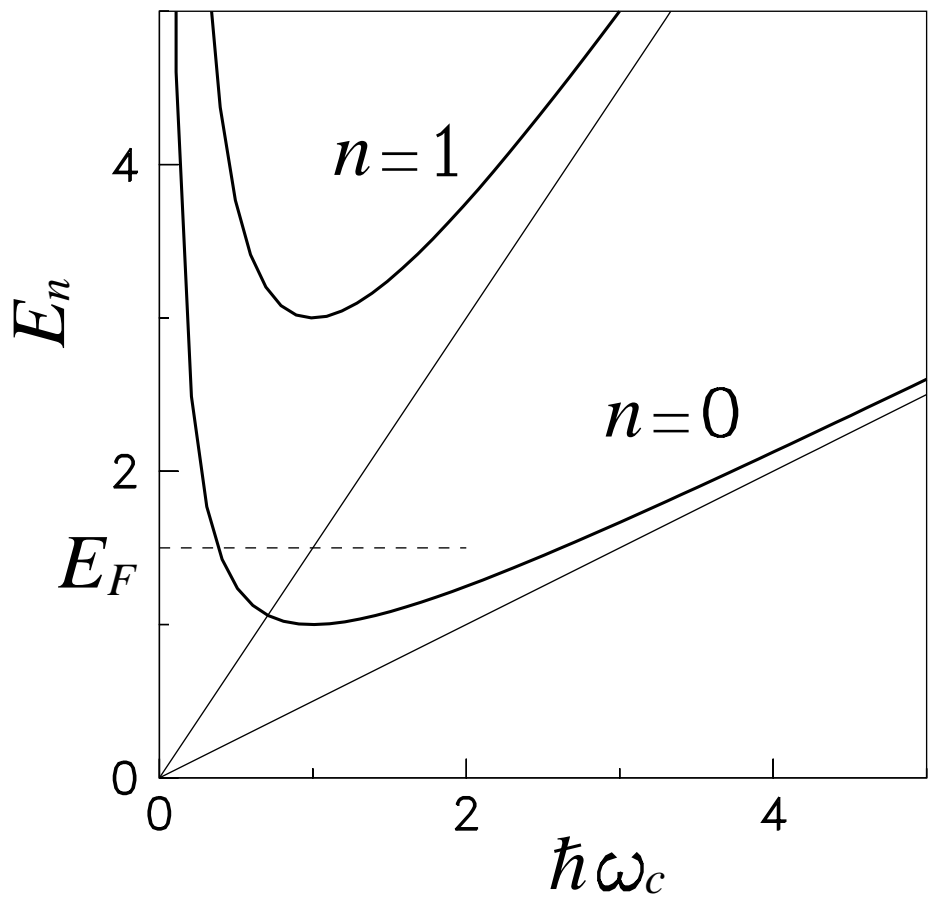
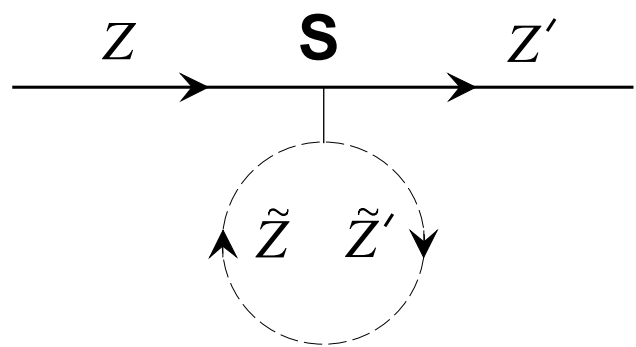
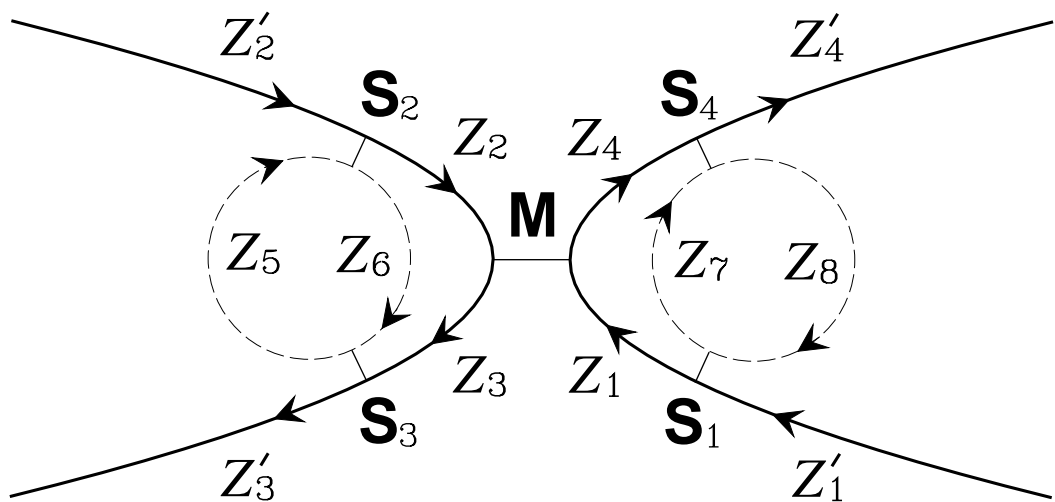
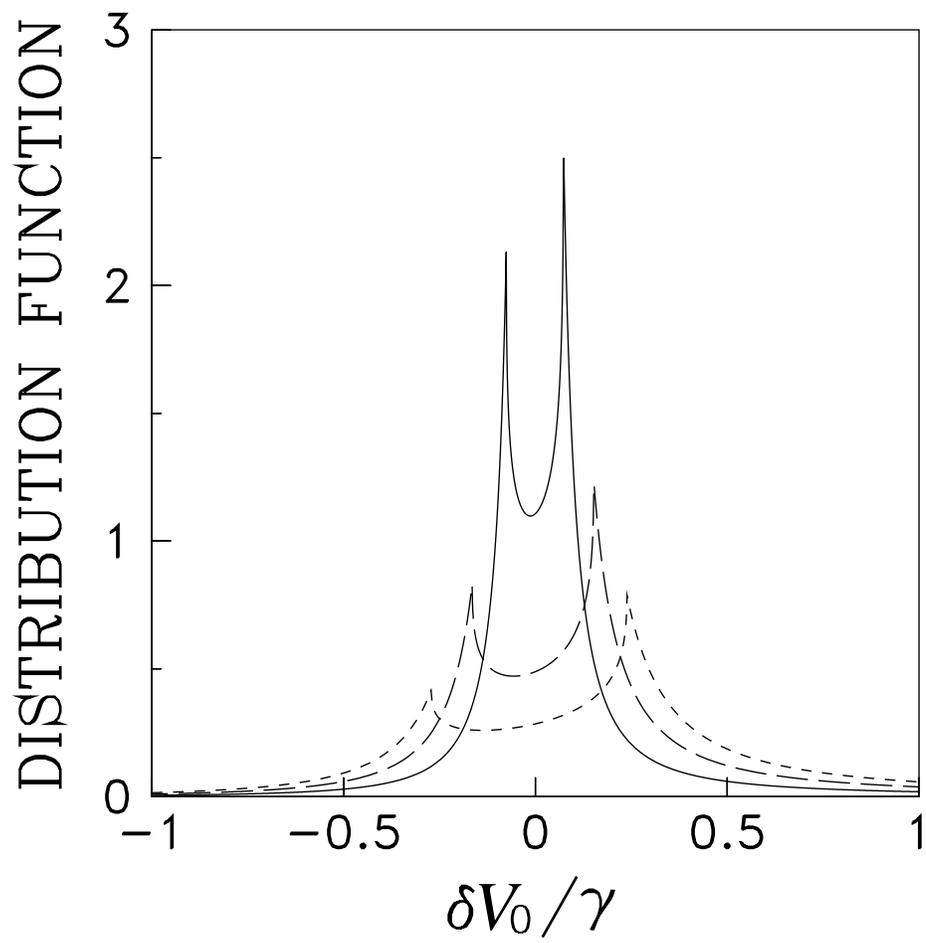
\end{document}